# Voltage-controlled transmission in a dielectric slab doped with the quantum dot molecules

**Lida Ebrahimi Zohravi and Mohammad Mahmoudi**

Department of Physics, University of Zanjan, University Blvd, 45371-38791, Zanjan, Iran

E-mail: mahmoudi@znu.ac.ir (Mohammad Mahmoudi)

Tel.: +98-241-5152521

Fax: +98-241-5152264

**Abstract**:
Transmission and reflection of an electromagnetic pulse through a dielectric slab doped with the quantum dot molecules is investigated. It is shown that the transmitted and reflected pulses depend on the inter-dot tunneling effect and can be controlled by applying a gate voltage.

## 1. Introduction

The reflected and transmitted pulses of an electromagnetic field from a dielectric slab were investigated [1]. Recently the phase control of the group velocity was studied in a dielectric slab doped with three-level ladder-type atoms [2]. Moreover, the coherent control of transmission and group velocity in one-dimensional photonic crystal was also introduced [3].

On the other hand, the quantum dots are interesting semiconductor devices which have tunable electronic and optical properties. A quantum dot (QD) is a semiconductor nanostructure that confines the motion of conduction band electrons in all three spatial directions and then electrons and holes can occupy only set of discrete energies. Theory of quantum coherence phenomena in the semiconductor QD is investigated [4] and coherent phenomena in ensembles of QDs have also been observed [5-7]. A QD molecule is a two-QD (the left and right one) coupled system, leading to the formation of coherent electronic states via inter-dot tunneling effect. The inter-dot tunneling effect can be controlled by a gate voltage [8]. The QD molecules have been extensively studied and it was shown that the optical properties of the system depend on the inter-dot tunneling effect [9-10]. Recently the voltage-controlled inter-dot tunneling effect has been used for controlling the optical bistability [9], entanglement [11] and light propagation [10].

In this letter, we consider a dielectric slab doped with the quantum dot molecules and investigate the reflected and transmitted pulses through the dielectric slab. It is shown that the transmission of the electromagnetic pulse can be controlled via a gate voltage. Moreover, the controllable transmission with a zero reflectivity is obtained for the slab.

## 2. Model and Solutions

We consider a weakly absorbing and nonmagnetic slap which is extended from $z = 0$ to $z = d$ as depicted in figure 1. A light pulse is normally incident on the slap, located in vacuum, with complex relative permittivity $\varepsilon(\omega_p) = \varepsilon_r + i\varepsilon_i$ where $\varepsilon_r$ and $\varepsilon_i$ represent the dispersion and absorption properties of the slab, respectively. The transfer matrix for electric and magnetic components of a monochromatic wave with frequency $\omega_p$ through the slab can be written as [1, 12]

$$\begin{pmatrix} \cos[kd] & i\frac{1}{n(\omega_p)}\sin[kd] \\ i n(\omega_p)\sin[kd] & \cos[kd] \end{pmatrix}, \tag{1}$$

where $n(\omega_p) = \sqrt{\varepsilon(\omega_p)}$ shows the refractive index of the slab. We assume a dielectric slab which is doped with quantum dot molecules and then the complex relative permittivity is given by two parts

$$\varepsilon(\omega_p) = \varepsilon_b + \chi(\omega_p), \tag{2}$$

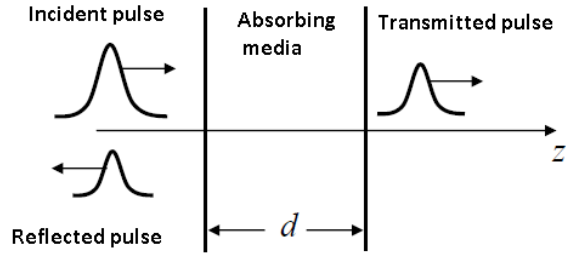

Figure 1

where $\varepsilon_b = n_b^2$ is the background dielectric function and $\chi(\omega_p)$ represent the susceptibility of quantum dot molecules. Using the transfer matrix method, the reflection and transmission coefficients of the monochromatic wave can be written as [12]

$$r(\omega_p) = \frac{-\frac{i}{2}\left(\frac{1}{\sqrt{\varepsilon}} - \sqrt{\varepsilon}\right)\sin(kd)}{\cos(kd) - \frac{i}{2}\left(\frac{1}{\sqrt{\varepsilon}} + \sqrt{\varepsilon}\right)\sin(kd)},\qquad (3\text{-a})$$

$$t(\omega_p) = \frac{1}{\cos(kd) - \frac{i}{2}\left(\frac{1}{\sqrt{\varepsilon}} + \sqrt{\varepsilon}\right)\sin(kd)}.\qquad (3\text{-b})$$

These equations show that the susceptibility of the doped elements has a major role in determination of reflectivity and transmission of a light pulse through the slab. Moreover, these coefficients depend on the thickness of the slab. The resonance condition in a passive slab happens at $d = 2m(\lambda_0/4\sqrt{\varepsilon_b})$, while the off-resonance condition is established at $d = (2m+1)(\lambda_0/4\sqrt{\varepsilon_b})$.

Let us consider a QD molecule consisting of a two-QD (the left and right one) system coupled by inter-dot tunneling. Such a QD molecule can be fabricated using self-assembled dot growth technology [8]. As a realistic example, the asymmetric QD molecules have been detected in double layer InAs/GaAs structures [13]. Two levels $|0\rangle$ and $|1\rangle$ are the lower valance and upper conducting band levels of the left QD, respectively. Level $|2\rangle$ is the excited conducting level of the second QD, as shown in figure 2(a). It is assumed that, the energy difference of two lower levels as well as that of two excited states is large, and then their tunneling couplings can be ignored. By applying a gate voltage, the level $|2\rangle$ gets closer to the level $|1\rangle$, while the valance band levels have still a high energy difference. The system configuration after applying the gate voltage is shown in figure 2(b). A weak probe field of frequency $\omega_p$ with Rabi-frequency $\Omega_p$ applies to the transition $|1\rangle \to |0\rangle$ (figure 2(c)). The Hamiltonian of the system is given by

$$H_{int} = \begin{pmatrix} \frac{1}{2}\delta & \Omega_p & 0 \\ \Omega_p & -\frac{1}{2}\delta & T_e \\ 0 & T_e & -\frac{1}{2}\delta - \omega_{12} \end{pmatrix},\qquad (4)$$

where $\delta = \omega_p - \omega_{10}$ is the probe detuning from the exact resonance, and $\omega_{1j}(j=0,2)$ is the central frequency of transition $|1\rangle - |j\rangle$. The parameter $T_e$ is also the electron tunneling matrix element. The probe field excites one electron from valance to conducting band in left QD that can tunnel to the right one.

In weak probe field approximation, i. e., $\Omega_p \ll \Gamma_{10}$, the probe susceptibility is given by [10]

$$\chi(\omega_p) \propto \frac{\rho_{10}}{\Omega_p} = \frac{(\delta + i\Gamma_{20})}{\left[(\Gamma_{10} - i\delta)(\Gamma_{20} - i\delta) + T_e^2\right]},\qquad (5)$$

where $\Gamma_{10}$ and $\Gamma_{20}$ are dephasing broadening of the corresponding transitions.

Equation (5) shows that the imaginary part is strictly positive doublet absorption and the maximums of the doublet absorption are located at $\delta_{min} \cong \pm T_e$.

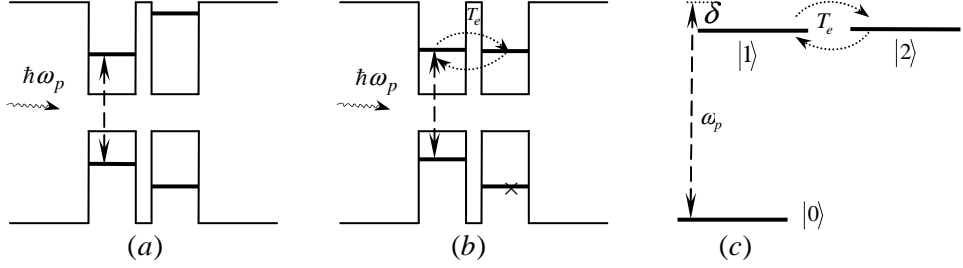

Figure 2

The imaginary parts of $\chi(\omega_p)$ around $\delta = 0$ is given by

$$\text{Im}[\chi(\omega_p)] = \frac{\Gamma_{20}}{T_e^2 + \Gamma_{10}\Gamma_{20}}, \tag{6}$$

which introduce a tunneling induced transparency like window for different values of inter-dot tunneling matrix element. The maximum value of absorption ($\cong 1/\Gamma_{10}$) is obtained for $T_e \ll \Gamma_{20}$ and the minimum value ($\Gamma_{20}/T_e^2$) is obtained for $T_e \gg \Gamma_{20}$.

The half width of the central dip around $\delta = 0$, for $T_e, \Gamma_{10} \gg \Gamma_{20}$ is also determined by

$$w \cong \sqrt{\frac{1}{2}\left(2T_e^2 + \Gamma_{10}^2 - \Gamma_{10}\sqrt{4T_e^2 + \Gamma_{10}^2}\right)} \tag{7}$$

which introduce a inter-dot tunneling broadening in absorption spectrum.

## 3. Results and discussion

Now we consider a dielectric slab doped with the quantum dot molecules and investigate the transmission and reflection of light pulse though the slab. In our calculation, times are given in units of the Planck constant, so that the energy of $1 meV$ corresponds to a frequency 242 GHz. In figure 3, we plot the transmission (a) and reflection (b) of light pulse versus detuning for different values of inter-dot tunneling parameters. Using parameters are $\Gamma_{10} = 5.54 meV$, $\Gamma_{20} = 0.005 meV$, $\omega_{12} = 0.0$, $\Omega_p = 0.001 meV$, $m = 50$ and $T_e = 0.1 meV$ (thin-solid), $0.5 meV$ (dashed), $1 meV$ (dotted), $2 meV$ (dashed-dotted), $2.5 meV$ (thick-solid).

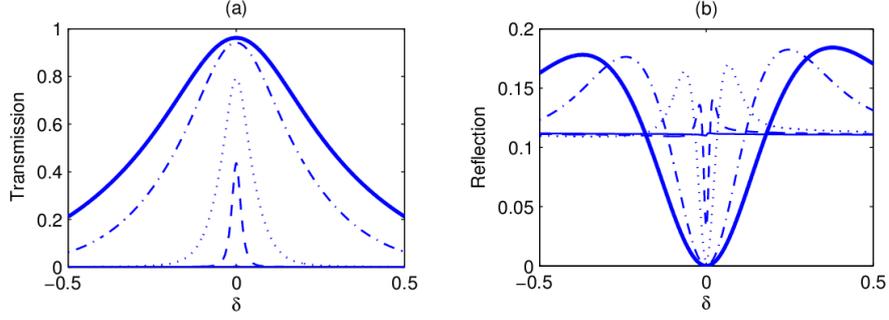

Figure 3

An investigation on figure 3(a) shows that the maximum transmission depends on the inter-dot tunneling parameter. For the small values of $T_e$, the magnitude of maximum transmission around zero probe detuning is negligible. However, by increasing the inter-dot tunneling via an applied gate voltage, the transmission increase and even the slab becomes transparent. Then the inter-dot tunneling effect induces a wide range of transmission for a slab doped by QD molecules. Note that the width of transparent window ($\approx 1 meV$) is approximately equal to the transparent width of the free QD molecules which is given by equation (7). The results of figure 3(b) show that increasing the inter-dot tunneling parameter applies a zero reflection for the slab and it can be used as an anti-reflected device; however it does not induce a wide range of variation for reflectivity of the slab.

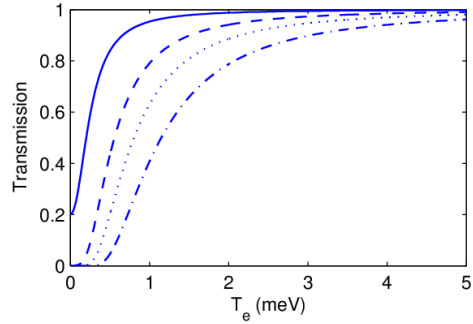

Figure 4

Other important parameter is the thickness of the slab. Figure 4 displays the behavior of maximum values of transmission versus the inter-dot tunneling parameter for different thickness of the slab. Using parameters are $\delta = 0$, $m = 10$ (solid), $50$ (dashed), $100$ (dotted), $200$ (dash-dotted). Other parameters are same as figure 3. It can be found that the thin slab, i.e. $m = 10$, cannot cover all possible values of transmission, but the problem is solved by choosing the thicker slab, i.e. $m = 50$ or $m = 100$.

Finally, we plot the transmission versus thickness of the slab for different values of inter-dot tunneling. Using parameters are $T_e = 0.1 meV$ (thin-solid), $0.5 meV$ (dashed), $1 meV$ (dotted), $2meV$ (dashed-dotted), $2.5meV$ (thick-solid). It can be seen that the increasing the thickness of the slab reduces the transmission, however the lope of reduction is steep for small values of inter-dot tunneling parameters.

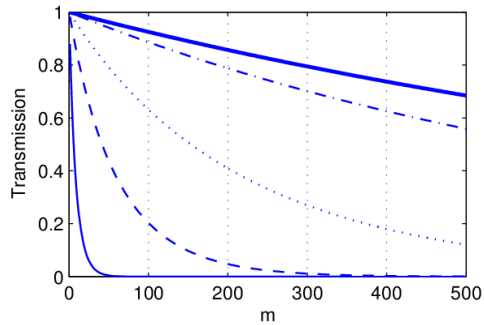

Figure 5

## 4. Conclusion

We showed that the optical properties of the QD molecules inside the slab are completely different from the gas systems. Moreover, it was demonstrated that by changing the applied gate voltage, the different values of transmission from zero up to one can be generated in a dielectric slab doped with the QD molecules.

**Figures caption**

Figure 1. Schematic of the weakly absorbing dielectric slab.

Figure 2. Band diagram of QD molecule interacting with a tunable weak probe field before (a) and after (b) applying the gate voltage. The dashed arrow shows the probe field. (c) Proposed energy levels scheme.

Figure 3. Transmission (a) and reflection (b) of light pulse versus detuning for different values of inter-dot tunneling parameters. Using parameters are $m = 50$, $\omega_{12} = 0.0$, $\Gamma_{10} = 5.54 meV$, $\Gamma_{20} = 0.005 meV$, , $\Omega_p = 0.001 meV$ and $T_e = 0.1$(thin-solid), $0.5 meV$ (dashed), $1 meV$ (dotted), $2 meV$ (dashed-dotted), $2.5 meV$ (thick-solid).

Figure 4. The behavior of maximum values of transmission versus the inter-dot tunneling parameter for different thickness of the slab. Using parameters are $\delta = 0$, $m = 10$ (solid), 50 (dashed), 100 (dotted), 200 (dash-dotted). Other parameters are same as figure 3.

Figure 5. Transmission versus thickness of the slab for different values of inter-dot tunneling. Using parameters are $T_e = 0.1$(thin-solid), $0.5 meV$ (dashed), $1 meV$ (dotted), $2 meV$ (dashed-dotted), $2.5 meV$ (thick-solid). Other parameters are same as figure 3.